\documentclass[aip,cha,reprint]{revtex4-1}

\usepackage{graphicx}  \usepackage[english]{babel}
\usepackage{pifont}
\usepackage{amssymb, amsfonts, amsmath}  \usepackage{units}
\usepackage{lineno}
\usepackage{hyperref}

\begin{document}
\draft
\title{\large A few points suffice: Efficient large-scale computation of   brain voxel-wise functional connectomes from a sparse spatio-temporal point-process}
\author{Enzo Tagliazucchi}
\affiliation{Department of Neurology and Brain Imaging Center, Goethe University, Frankfurt am 
Main, Germany}
\author{Helmut Laufs}
\affiliation{Department of Neurology and Brain Imaging Center, Goethe University, Frankfurt am 
Main, Germany}
\affiliation{Department of Neurology, University Hospital Schleswig Holstein, Kiel, Germany.}
\author{Dante R. Chialvo}
 \affiliation{Consejo Nacional de Investigaciones Cientificas y Tecnologicas (CONICET), 
Buenos Aires, Argentina.}
\date{\today}

\begin{abstract}
Large efforts are currently under way to systematically map functional connectivity between all pairs of millimeter-scale brain regions using big volumes of neuroimaging data. Functional magnetic resonance imaging (fMRI) can produce these functional connectomes, however, large amounts of data and lengthy computation times add important overhead to this task. Previous work has demonstrated that fMRI data admits a sparse representation in the form a discrete point-process containing sufficient information for the efficient estimation of functional connectivity between all pairs of voxels. In this work we validate this method, by replicating results obtained with standard whole-brain voxel-wise linear correlation matrices in two datasets. In the first one (n=71) we study the changes in node strength (a measure of network centrality) during deep sleep. The second is a large database (n=1147) of subjects in which we look at the age-related reorganization of the voxel-wise network of functional connections.  In both cases it is shown that the proposed method compares well with standard techniques, despite requiring of the order of 1\% of the original fMRI time series. Overall, these results demonstrate that the proposed approach allows efficient fMRI data compression and a subsequent reduction of computation times.  

\end{abstract}
\maketitle
 
 \section{Introduction}

The human brain comprises an interconnected network of cortical and sub-cortical regions globally linked by long-range tracts of anatomical connections. The mapping of such connections at a particular spatial scale (dubbed connectome in contemporary neuroscience; Sporns et al., 2004; Sporns, 2011) is an important ingredient in the process of understanding how the human brain can perform diverse cognitive functions.  Furthermore, many neurological and psychiatric diseases can be understood in terms of deviations from a healthy connectome (Fox and Greicius, 2010; Kelly et al., 2012).

Advances in neuroimaging methods, such as Diffusion Tensor Imaging (DTI) and Diffusion Spectrum Imaging (DSI) allow the in vivo mapping of the human structural connectome at a large-scale (Hagmann et al., 2008). Functional Magnetic Resonance Imaging (fMRI) allows for a functional counterpart of the anatomical connectome, a notion first introduced a decade ago (Eguiluz et al., 2005, Salvador et al., 2005, Sporns et al., 2004) by computing the statistical covariance between all pairs of Blood Oxygen Level Dependent (BOLD) signals. This functional connectome contains dynamical information on how all pairs of regions (at a certain spatial scale) relate collectively with each other. 
 
These two approaches are being applied by international coordinated efforts to systematically map connectomes in very large populations of subjects and at the highest temporal and spatial resolution currently available (see for instance Biswal et al., 2010; Smith et al., 2013; Van Essen et al., 2013). These efforts will eventually lead to the availability of large-scale databases useful to account for potential inter-subject variability caused by different demographical variables, as well as to reduce the harmful effect of noise and artifacts through massive averaging. 

These collaborative efforts need to be paralleled by methodological developments allowing efficiently storage of data and extraction of relevant information (for a recent review on the equivalent problem for electrophysiological recordings, see Cunningham and Yu, 2014). Common strategies are based on averaging BOLD signals over brain parcellations comprising extended regions, thus reducing the dimensionality of the problem as well as the number of required computations. However, there are many problems inherent to this approach. First, all detail of the functional connectome inside each region of the parcellation is lost. Second, partitions are usually arbitrary and therefore might sub-divide a functionally coherent region into many regions. Different studies have addressed how the properties of parcellation-based networks can change depending on region selection (Wang et al., 2009; Zalesky et al., 2010). Third, efforts to increase the spatial resolution of fMRI sequences are pointless if data will be down-sampled after acquisition by averaging BOLD signals inside a small number of regions in a parcellation.

The objective of this paper is to show how a very sparse representation of brain activity, namely a discrete spatio-temporal point-process, is able to efficiently estimate the whole brain voxel-wise functional connectome. This point-process is derived from the times at which the BOLD signals reach some maximum level of activity, either by detecting crossings of an arbitrary threshold, or by the identification of local peaks, i.e. the point-process comprises large amplitude events in the data. 

It has been previously shown that this method suffices to reproduce large-scale patterns of coordinated activity (Tagliazucchi et al. 2011, Tagliazucchi et al., 2012A) (Resting State Networks – RSN; Beckmann et al., 2005) and is essentially identical to the de-convolution of the signals as a series of discrete impulse functions (Petridou et al., 2013). Here we show that, in fact, not only the discrete set of RSN but also all bivariate relationships between signals (i.e. whole-brain correlation matrices) can be estimated from the point-process. Additionally, we propose a framework to do so efficiently. This sparse representation of fMRI data can reduce computation times and, more drastically, memory requirements, thus being relevant to projects involving a large number of subjects scanned with a high spatial resolution.

\section{MATERIALS AND METHODS}

We will first describe all steps of the proposed method and then introduce different datasets used for validation as well as to show possible applications. The general procedure followed to estimate correlation networks via the point-process analysis is graphically outlined in Fig. 1.

\textbf{Voxel-wise correlation matrix}

Consider a fMRI measurement consisting of $N$ voxels and $T$ volumes, represented as $F_n(t)$, with $1 \leq n \leq N$ and $1 \leq t \leq T$. Thus $F_n(t)$ represents the BOLD signal at voxel $n$ and time $t$. The common definition of voxel-wise correlation network (Eguiluz et al., 2005) is as follows,

\begin{equation}
R_{ij} = \frac{<(F_i - <F_i>)(F_j - <F_j>)>}{\sigma(F_i) \sigma(F_j)}
\end{equation}

where $<F_i>$ and $\sigma(F_i)$ represent the mean value and the standard deviation of the BOLD signal at the voxel $i$, respectively. Note that according to this definition, Eq. 1 must  be evaluated a total of $\frac{N(N-1)}{2}$ times, although not serially in efficient implementations. Often these calculations are used to define functional connectivity networks which in turn allow for further analysis of the resulting graphs.

\textbf{Constructing the point-process}

The approach here proposed starts converting the BOLD signal at every voxel into its z-score, $\bar{F}_i = \frac{F_i - <F_i>}{\sigma(F_i)}$. .  This is done under the assumption that, according to our formalism, the absolute amplitude of the BOLD signal carries less information than its temporal evolution.  To define the point-process, the a priori arbitrary threshold $\gamma$ is selected and the spatio-temporal process $PP_i(t)$ is defined as follows:

\begin{equation}
    PP_i(t) = \begin{cases}
               1              & F_i(t) < \gamma \text{  and  } F_i(t+1) > \gamma \\
               0               & \text{otherwise} 
           \end{cases}
\end{equation}

This point-process was introduced in a previous publication (Tagliazucchi et al., 2012A) where we showed that it suffices to replicate the topographical features of the major canonical Resting State Networks (RSN), even though for most values of $t$ and $i$, $PP_i(t)$ will be zero (indeed, taking $\gamma=1$, for a signal of T=240 on average the point-process is non-zero for 15$\pm$3 time points, see Tagliazucchi et al., 2012A).

Alternatively, $PP_i(t)$ can be defined by the (high amplitude) local peaks of the BOLD signal. For this, BOLD signals are also converted to z-scores and all sufficiently large peaks (for instance, those above an arbitrary threshold) are the points represented in $PP_i(t)$. The formal definition is as follows,

\begin{equation}
    PP_i(t) = \begin{cases}
               1              & F_i(t) > F_i(t-1) \text{  and  }   F_i(t) > F_i(t+1)  \\
               0               & \text{otherwise} 
           \end{cases}
\end{equation}

Although formally both methods are justified, it will be shown later that either definition of the point-process leads to similar results.

\textbf{Estimating correlations from the point-process}

After converting $F_i(t)$ into $PP_i(t)$ we introduce the following framework to generalize the methods introduced in Tagliazucchi et al., 2012A, from the estimation of seed based correlations to the efficient computation of all pairs of correlations between voxels. We first define the co-activation matrices $A_{ij}(t)$ as follows:

\begin{equation}
 A_{ij} = PP_i(t)PP_j(t)
\end{equation}

Note that according to this definition,  $A_{ij}(t)$ only has two possible values:  $A_{ij}(t)=1$ if at time $t$ the point-process is non-zero both at voxels $i$ and $j$, and $A_{ij}(t)=0$  otherwise.

The co-activation matrices defined in Eq. 4 can be used to estimate the functional connectivity between all pairs of voxels in the brain by performing a simple matrix addition. Two highly synchronized signals will cross the threshold γ together most of the time, thus a measure of coupling between the signals can be obtained by counting the number of times the signals crossed the threshold together. This is formalized simply by,

\begin{equation}
 C_{ij} = \sum_{t=0}^T A_{ij}(t) =  \sum_{t=0}^T PP_i(t)PP_j(t)
\end{equation}

In matrix notation, this can be succinctly summarized as $C = PP \cdot PP^T$, considering $PP$ as a matrix with voxels as rows and time as columns and containing the point-process. The matrix $C_{ij}$ contains in its $i,j$ entry the number of shared co-activations between BOLD signals at voxels $i$ and $j$.  Note that since all $A_{ij}$ are symmetrical matrices, then $C_{ij}$ is also symmetrical. Note also that $C_{ij}$  contain valuable information about instantaneous co-activations and as such, their analysis might be important to understand the temporal evolution of large-scale synchronization of brain regions (Tagliazucchi et al., 2012B, Hutchison et al., 2013).

The main issue with this matrix as a measure of functional connectivity is that it is not normalized, therefore there is no way to directly decide (for instance) if a perfect synchronization between signals has been reached. The correct normalization for this matrix is as follows,

\begin{equation}
\bar{C}_{ij} = \frac{C_{ij}}{\max(\sum_{t=0}^T PP_i, \sum_{t=0}^T PP_i)} = \frac{C_{ij}}{\max(C_{ii}, C_{ij})}
\end{equation}

This definition of $\bar{C}_{ij}$ is reasonable since $C_{ij}$ achieves its highest possible value if all threshold crossings are also shared between both voxels. However, one voxel could have all its threshold crossings in common with the other, whereas the opposite might not be true (since the other voxel could have a larger number of crossings in total, this is the case only if $C_{ii} \neq C_{jj}$), thus normalizing using the maximum between the number of crossings at both voxels is required. Also, $\bar{C}_{ij}$ is symmetrical with this normalization.

The normalization presented in Eq. 6 requires the maximum value between the number of threshold crossings at all pairs of voxels. If normalization is needed, then a more efficient approximate solution is to divide by the number of threshold crossings without taking the maximum value, for instance, across rows or columns of the matrix, and then symmetrizing  (if needed) the result by averaging with the transpose:

\begin{equation}
\bar{C}_{ij} = \frac{1}{2} \left[ \frac{C_{ij}}{C_{ii}} + \frac{C_{ij}^T}{C_{ii}} \right]
\end{equation}

Note that $\frac{C_{ij}}{C_{ii}}$  deviates from a symmetrical matrix only in the case of different number of threshold crossing between voxels ($C_{ii} \neq C_{jj}$). Note also that normalization might not be necessary if comparing fixed-length recordings between two populations, under the assumption that the rate of events in the point-process is not different between groups.

For the computation of $\bar{C}_{ij}$ all steps can be performed efficiently in vectorized form in any language with matrix manipulation capabilities (for instance, MATLAB or Python with NumPy), in particular, after constructing the point-process in Eqs. 2 and 3, the operations involved consist of a single matrix multiplication (Eqs. 4 and 5), multiplication by scalars and matrix symmetrization (Eq. 7).  In this work, all computations were performed using a 8 core CPU running at 2400 MHz with a total of 128 GB built-in memory.

\textbf{Observables from whole brain voxel-wise correlations used for method validation}

The number of connections derived in a voxel-wise analysis complicates easy visualization of networks and their changes across conditions. Thus, in the many applications of functional connectomes found in the literature, rarely whole-brain voxel-wise networks are directly visualized. Instead, lower-dimensional observables are to be derived, which are easy to visualize as 3D maps overlaid on brain anatomy. One possible choice is to assess measures of network centrality, this is, “how important” nodes are in the network, thus collapsing all connections attached to a node into a single number. A straightforward definition in a weighted network is the strength (Barthelemy et al., 2005), defined as:

\begin{equation}
S_i = \sum_{j=1}^N R_{ij}
\end{equation}

In the present case, using the point-process to estimate correlations, $R_{ij}$ is replaced by $\bar{C}_{ij}$. Nodes with the highest strength values are termed hubs and their reorganization has been repeatedly linked to different brain pathologies (Crossley et al., 2014), such as coma (Achard et al., 2009) and Alzheimer (Buckner et al., 2009).

Note that the evaluation of Eq. 8 requires the whole brain correlation network. In the case of a voxel-wise network, centrality of nodes (i.e. voxels) can be easily visualized as a 3D map overlaid on an anatomical image.

Another observable employed for validation of our method is the interhemispheric or homotopic connectivity. This is defined as the correlation between the BOLD signal of every voxel and the contralateral voxel. Interhemispheric connectivity is in particular useful to quantify re-organization of functional connectomes for which left-right asymmetries are expected (as in the case of aging, see Dolcos et al., 2002).

\textbf{Datasets}

To demonstrate the validity of the proposal two different datasets from previously published studies will be used. The first dataset comprises fMRI recordings from the 1000 Functional Connectomes database and the second dataset comprises recordings from a recently published study in which combined EEG, EMG, fMRI and physiological data were obtained from 71 subjects.

The Connectome dataset was downloaded from the 1000 Functional Connectome Project online database (\url{http://fcon_1000.projects.nitrc.org}). Demographics, scanning parameters and experimental conditions are described in the database website as well as in Tagliazucchi and Laufs, 2014. Only epochs of wakefulness were employed in the present analysis. For more information on sleep vs. wakefulness classification in this dataset see Tagliazucchi and Laufs, 2014.

Data from a previously published study (Tagliazucchi et al., 2014) was used for the sleep dataset. A total of 71 subjects were selected from a larger dataset on the basis of successful EEG, EMG, fMRI and physiological data recording and quality (written informed consent, approval by the local ethics committee). All subjects were scanned during the evening and instructed to close their eyes and lie still and relaxed. A group of 55 subjects was formed out of the original dataset of 71 subjects (by excluding subjects who did not fall asleep). Hypnograms (obtained after expert sleep staging based on AASM rules) were scanned for contiguous epochs of wakefulness, N1, N2 and N3 sleep lasting 250 volumes (approximately 2 minutes), resulting in 84 epochs of wakefulness, 16 epochs of N1 sleep, 19 epochs of N2 sleep and 20 epochs of N3 sleep.

EEG was recorded via a cap (modified BrainCapMR, Easycap, Herrsching, Germany)  during fMRI acquisition (1505 volumes of T2$^*$-weighted echo planar images, TR/TE = 2080 ms/30 ms, matrix 64$\times$64, voxel size 3$\times$3$\times$2 mm$^3$, distance factor 50\%; FOV 192 mm$^2$) at 3 T (Siemens Trio, Erlangen, Germany) with an optimized polysomnographic setting (chin and tibial EMG, ECG, EOG recorded bipolarly [sampling rate 5 kHz, low pass filter 1 kHz], 30 EEG channels recorded with FCz as the reference [sampling rate 5 kHz, low pass filter 250 Hz], and pulse oxymetry, respiration recorded via sensors from the Trio [sampling rate 50 Hz]) and MR scanner compatible devices (BrainAmp MR+, BrainAmp ExG; Brain Products, Gilching, Germany).

MRI and pulse artifact correction were performed based on the average artifact subtraction (AAS) method (Allen et al., 1998) as implemented in Vision Analyzer2 (Brain Products, Germany) followed by objective (CBC parameters, Vision Analyzer) ICA-based rejection of residual artifact-laden components after AAS resulting in EEG with a sampling rate of 250 Hz. Good quality EEG was obtained, which allowed sleep staging by an expert according to the AASM criteria (AASM, 2007).

\textbf{fMRI preprocessing}

Using Statistical Parametric Mapping (SPM8) EPI data were realigned, normalized (MNI space) and spatially smoothed (Gaussian kernel, 8 mm full width at half maximum). Data was band-pass filtered in the range 0.01-0.1 Hz using a sixth order Butterworth filter. The same procedure was applied to the sleep dataset and to the 1000 Functional Connectomes dataset.

\section{RESULTS}

\textbf{Correlation between $\bar{C}_{ij}$ and $R_{ij}$}

We obtained the point-process for both datasets following the procedure illustrated in Fig. 1 and in the methods section. In the case of the 1000 Functional Connectomes dataset we repeated calculations both for voxel-wise networks and for networks based on time series extracted from the AAL template. Using this data, we first evaluated the similitude in the estimation of the connectivity matrix by both methods (point-process analysis and linear correlations) as a function of the threshold $\gamma$ used to define the point-process (see Eq. 2). Results are shown in Fig. 2 (left) for the average correlation between connectivity networks estimated by both methods as a function of $\gamma$. Correlations peaked at 0.6 and were highest for γ $\approx$ 0.7. The histogram of all 1147 correlations obtained at γ $\approx$ 1 (Fig. 2, center) revealed a sharp peak around the mean value. The plot of the entries of the estimated correlation (values of $\bar{C}_{ij}$) and the linear correlation (entries of $R_{ij}$) is shown in Fig. 2 (right). A monotonously increasing relationship was present between both quantities, even though the functional dependency between them was not linear. For low linear correlation values, the point-process co-activation increased slowly and did so more quickly for larger linear correlation values. 

We compared the time performance of computing voxel-wise functional connectivity matrices using the proposed point-process based method with standard linear correlations. In Fig. 2B, left, the percentage of the time required using linear correlations (\textbf{corrcoef.m} MATLAB function, average time 131.48 s.)  was plotted as a function of the threshold. At every threshold value a total of 100 iterations were performed for a single subject and results were then averaged. For thresholds larger than approximately 1 standard deviation, the point-process based method outperformed the standard computation, with performance becoming increasingly better as the threshold was increased and less points were included in the analysis. In Fig. 2B (right) we plot the percentage of data points retained after conversion to the point-process.  Even for the smallest threshold values, only about 6\% of the data was retained. Thus, this very sparse representation of fMRI data contained sufficient information to capture all the aforementioned differences during deep sleep and in the 1000 Functional Connectomes dataset, but requiring a small fraction of the original time series.

To gauge the usefulness of our approach in a real setting, we computed the cumulative time and space required to process (i.e. obtain whole-brain voxel-wise connectivity matrices) and store 1000 subjects extracted from the Functional Connectomes dataset. Results are shown in Fig. 2C. An un-normalized point-process with threshold of $\gamma$=1 resulted in a reduction from a total of $\approx$30 hours to $\approx$19 hours. In terms of space required to store the data, the point-process representation resulted in a decrease from $\approx$19 TB to $\approx$5 TB (order of magnitude reduction).

\textbf{Strength maps}
 
To compare results obtained by both methods, we applied them to derive the strength maps (Eq. 8) from the estimated whole brain voxel-wise correlations in the sleep dataset and to reveal changes between wakefulness and deep sleep. A total of 20 epochs of deep sleep and 84 epochs of wakefulness could be extracted (all epochs lasting 2 minutes). After deriving the correlation networks, Eq. 8 was applied to obtain the voxel-wise spatial distribution of strengths. Results for the contrast wakefulness $>$ deep sleep are shown in Fig. 3A, both for normalized and un-normalized co-activation matrices, as well as for the point-process derived from BOLD signal peaks instead of threshold crossings. Spatial patterns of decreased strength in deep sleep (comprising frontal, cingulate, primary visual, motor and auditory cortices) were captured equally well by both methods, as well as by the peak-based point-process. In particular, since fixed epoch lengths were used (250 volumes) results were reproduced with and without normalization of connectivity matrices as derived from the point-process.  This similitude can also be seen in Fig. 3B, in which a joint 3D rendering of both maps shows their spatial agreement. The main plots in Fig. 3C show node strength values at all voxels computed using the point-process method (entries of $C_{ij}$) vs. those computed using linear correlations (entries of $R_{ij}$). The functional dependency was clearly monotonously increasing on average, both for wakefulness and sleep, although two individual epochs of sleep displayed an opposite trend.

We then studied changes in node strength in the 1000 Functional Connectomes dataset, in particular, we compared a group of subjects younger than 20 years with an older group of subjects older than 40 years. Results can be found in Fig. 4A. For both methods an increase of functional connectivity strength in the older group was observed, comprising a network of regions that included the right parietal cortex, inferior frontal cortex, insula and the precentral and postcentral gyrus.

Driven by the asymmetry observed in the strength differences between age groups, and by the proposal that the right hemisphere shows accelerated functional decline with aging (Dolcos et al., 2002), we applied linear correlations and the point-process analysis to quantify interhemispheric or homotopic connectivity between groups and compare the respective values. Results are shown in Fig. 4B. Increased interhemispheric connectivity was observed for the older group of subjects by both methods, comprising areas in the parietal and temporal cortex, as well as in the precentral gyrus.

Finally, an additional calculation was performed to allow for further evaluation of our method. We regressed subject age vs. strength values in two regions of interest extracted from the analysis of young vs. older subjects (right Inferior Parietal Cortex - IPC, right and insular cortex). Strength values were obtained both from connectivity matrices obtained with linear correlations and with the point-process. Results are shown in Fig. 4C. The plots show a moderate increase in strength with age, which suddently increases for more mature subjects (age $>$ 40 years approximately). Spearman's rank correlation coefficients were higher for the strength values computed using the point-process.

\section{DISCUSSION}

We are witnessing in recent times how neuroscience, and in particular neuroimaging, is moving at a fast pace towards the accumulation and analysis of very large volumes of data. A number of international collaborations is aiming to break new ground in the scale and speed of data collection, including the 1000 Functional Connectomes Project, as well as the Human Connectome Project. These studies span hundreds of subjects scanned at high temporal resolution, resulting in datasets large enough to render broadband downloading from their Internet servers almost impractical.

While it is obvious that having large volumes of data reduces the negative effect of noise, artifacts and the relative importance of the mathematical models employed to analyze it (a position eloquently defended by Halevy et al. (2009) in their seminal article “The Unreasonable Effectiveness of Data”), it is also true that the handling of data might become prohibitive beyond a given size and efficient data compression (both for storage and transfer of information) is of paramount importance. In this line of thought, we have shown that the introduction of a sparse representation of fMRI datasets can reproduce findings obtained from full time series while keeping on the order of 1\% of the original data.

In the present paper we validated our method by first computing correlation between connectivity matrices as obtained by both methods over $>$ 1000 subjects in the Functional Connectomes dataset, as well as by comparing voxel-wise network strength (a measure of centrality computed from the voxel-wise network of functional connections) between wakefulness and deep sleep and between two age groups extracted from the 1000 Functional Connectomes dataset.  In this latter dataset we also obtained the distribution of voxel-wise inter-hemispheric connectivity. The maps of altered network strength in deep sleep and the age-dependent effect observed in the 1000 Functional Connectomes dataset are of biological relevance themselves, as we are not aware of prior reports of these results. Deep sleep resulted in a loss of connectivity across all voxels located in frontal and cingulate cortices, as well as in the primary auditory cortex (Heschl gyrus) and the thalamus. These are plausible correlates of reduced awareness (frontal and cingulate cortex) and loss of sensory engagement with the environment (primary auditory cortex and thalamus) resulting in increased arousal thresholds (Tagliazucchi et al., 2013). 

With respect to the two different age groups extracted from the 1000 Functional Connectomes database, regions central to working memory processes (inferior parietal and frontal cortices, prefrontal cortex) showed over-connectivity in the older group of subjects. The meaning of this result is less clear, especially in the light of reports showing an inverse relationship between seed-based functional connectivity and age (Sambataro et al., 2010). However, voxel-based strength maps do not require any a priori anatomical hypotheses (i.e. seed selection) and thus might be capable of capturing more global changes in connectivity, as opposed to the aforementioned approach. Interestingly, changes in the node strength values were mostly located in the right hemisphere. It has been noted by Dolcos and colleagues (Dolcos et al., 2002) that the right hemisphere shows a more marked decline with aging, a fact supported so far by evidence from working memory neuroimaging experiments. The changes observed by the authors were hypothesized to be of compensatory origin, which is compatible with the outcome of our analyses (increased overall connectivity in the right hemisphere of older subjects). Prompted by this, we also found differences in interhemispheric connectivity located in a set of regions overlapping with those involved with changes in node strength.

\textbf{Why few points suffice to reproduce functional connectomes?}

It is worthwhile to discuss the reasons underlying the effectiveness of our approach, since it might be surprising that such small fraction of the data suffices to capture all bivariate relationships between BOLD signals (functional connectome). 

From a signal processing perspective the answer is relatively straightforward: keeping large amplitude events increases the signal-to-noise ratio, since it discards low-amplitude activity containing a larger noise component. This “non-linear filtering” selectively amplifies the importance of those time points at which the signal amplitude becomes relatively large and therefore the signal-to-noise ratio increases. 

From a biological point of view, the challenge is to understand why the fMRI time series can be effectively represented as a train of discrete impulses, a view of BOLD time series also supported by studies performing blind de-convolution of spontaneous activity (Petridou et al., 2013). Electrophysiological experiments reveal that Local Field Potentials (LFP) are spatio-temporally distributed as power law avalanches (Beggs and Plenz, 2003): most frequently, spontaneous LFP increases span a limited spatial area, however, at certain (discrete) points in time, LFP might extend up to the size of the tissue under study (an event termed “avalanche”). If LFP avalanches are, indeed, distributed following a scale-free power law, then macroscopic events (i.e. in the centimeter scale) should be observed, which would be sufficient to elicit a measurable hemodynamic response (considering the correlation observed between LFP and BOLD signals, see Logothetis et al., 2001). Indeed, spatio-temporal avalanches of activity can also be observed with fMRI, following the same statistical laws as the electrophysiological avalanches (Tagliazucchi et al., 2012A). Large amplitude macroscopic LFP increases were reported in the monkey cortex (Thiagarajan et al., 2010) and termed “coherence potentials”. These large-scale events are also stereotypical (in the words of the authors, “much like action potentials at the single-cell level”) and thus fulfill all the theoretical requirements for the electrophysiological underpinnings of the events in the spatio-temporal fMRI point-process.

\textbf{Caveats and limitations}

Generally, this procedure should yield equivalent results for any dataset in which high amplitude events do not arise spuriously as artifacts or noise and represent important information in the data. From a neurophysiological perspective, the fulfillment of these conditions has been already demonstrated for BOLD time series by means of inverting the Hemodynamic Response Function (HRF) convolution of neuronal sources (de-convolution). As discussed in the previos section, Local field potentials (LFP) giving rise to metabolic changes reflected in the BOLD signal are temporally cluttered into avalanches of activity (Beggs and Plenz, 2003; Tagliazucchi et al., 2012A; Shriki et al., 2013), presumably underlying the high information content of BOLD signal high amplitude events. 

The main drawbacks of the proposed method are: 1) the non-linear relationship between linear correlation and its estimated value using the point-process (i.e. point-process co-activation, Fig. 2C) and 2) the slowing down of the computation time when following the normalization given by Eq. 6, unless properly optimized. With respect to the first concern, while not linear, the relationship is clearly monotonic and by extracting its functional form, connectivity estimated using the point-process can be properly normalized to have a linear co-variation with standard functional connectivity. This non-linear shape can be explained by the dismissal of low amplitude events in the point-process and their associated contributions to linear correlations. Therefore, correlations can increase faster than point-process co-activations, giving rise to the convex shape seen in Fig.  2A, right panel. The second concern  (normalization) does not affect the results unless performing comparisons between time series of different length,  thus having a different number of points. Normalizing by the length of the time series offers a solution to this issue.

\textbf{Related findings}

Given the relative novelty of the present approach, caution should be exercised concerning the interpretation of the results to avoid making exaggerated claims.  Nevertheless, is encouraging and reassuring to see a body of publications consistent with the main idea of the present paper. Indeed, since the first observation (Tagliazucchi et al., 2012A) that the timing of high-activity events in BOLD signals allows the reconstruction of major RSN, different research groups have reproduced and built on this result (Liu and Duyn, 2013; Liu et al., 2013; Amico et al., 2014). The analysis of spontaneous voxel co-activation is a natural continuation of functional connectivity studies: instead of asking whether two voxels are engaged in synchronized fluctuations over a relatively long period of time, the question is shifted to whether two voxels become jointly activated (i.e. present high activity above their baseline levels) and what are the timings and properties of these co-activations. Interestingly, it has been shown that co-activation patterns contain additional information not available to standard functional connectivity analyses (Liu et al., 2013) and has also been used to characterize the dynamics of different brain states (Amico et al., 2014). In the present report we show that the spatio-temporal point-process extracted from whole-brain BOLD signals suffices to estimate all pairs of functional connections (i.e. the functional connectomes) with reasonable accuracy (as demonstrated by its usefulness to capture differences in connectivity between brain states/groups of subjects) with a very small fraction of the data (on the order of 1\%), and thus can be taken as an equivalent (but sparser) representation of the data. We believe these results should prompt an in-depth exploration of high amplitude events in BOLD time series, in particular, their neural correlates and potential relationship to LFP neural avalanches, a signature of self-organized criticality in the human brain (Chialvo, 2010).

In conclusion, as fMRI datasets grow larger, tools to rapidly store, process and explore them become increasingly valuable. The present report validates a strategy defining a sparse representation of these complex four-dimensional datasets, which keeps only the timing of large BOLD events and thus allows for reasonable and efficient fMRI compression. This technique should empower standard desktop computers to store and analyze fMRI data coming from vast collaborative projects, thus adding value to this data.

\section*{Acknowledgements}
Work supported by CONICET (Argentina) and LOEWE Neuronale Koordination Forschungsschwerpunkt Frankfurt - NeFF (Germany).

\begin{figure*}[ht]
\centering \includegraphics[width=0.8\textwidth,clip=true,angle=0]{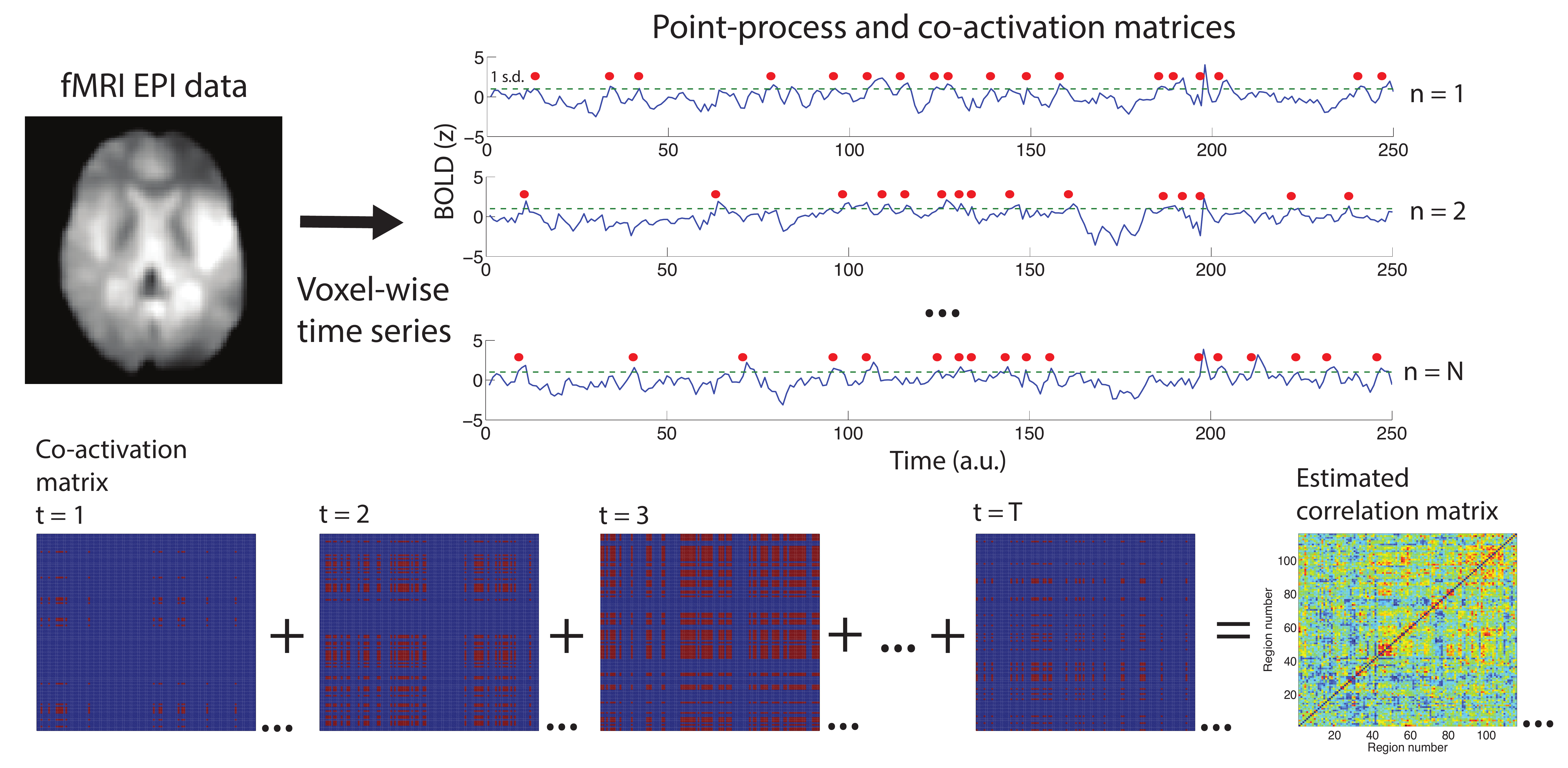} 
\caption{Procedure to construct the point-process and to estimate functional connectomes. For every voxel, signals are converted to z-scores and a discrete event marked after every threshold crossing (in this example the threshold was set to 1 standard deviation). For every volume a whole brain co-activation matrix is derived, and the sum of all co-activation matrices estimates the functional connectivity matrix or correlation matrix (only a fraction of the matrices are shown in this example).}
\end{figure*}

\begin{figure*}[ht]
\centering \includegraphics[width=0.7\textwidth]{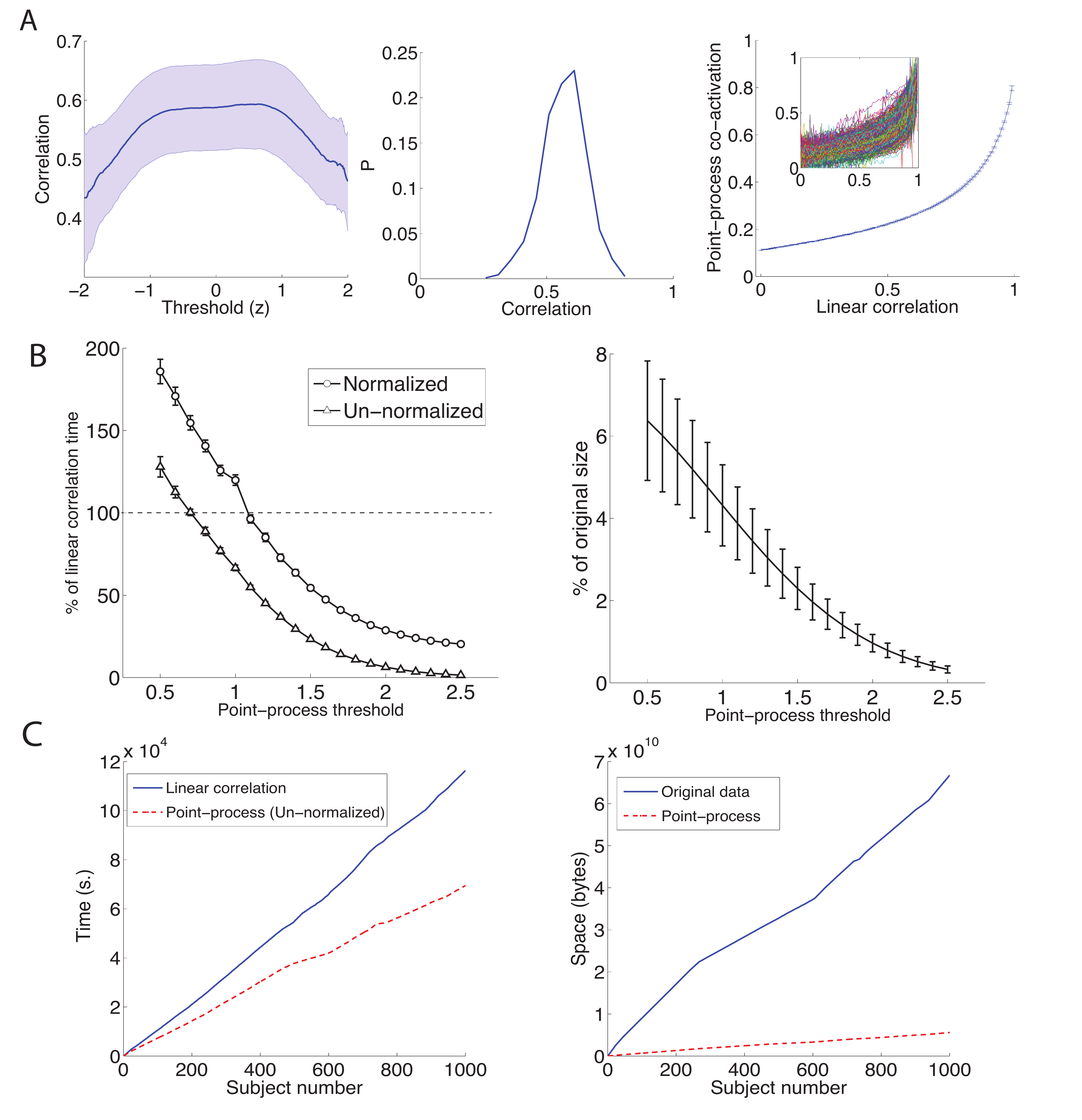} 
\caption{A. Left: Correlation between $R_{ij}$ and $\bar{C}_{ij}$ as function of the threshold ($\gamma$ in Eq. 2) (mean $\pm$ SEM). Connectivity wetworks were derived from 116 time series extracted from the AAL template in all subjects from the 1000 Functional Connectomes dataset (n=1147). Center: Histogram of all correlation values at $\gamma$=1. Right: Average (mean $\pm$ SEM) plot of the linear correlation coefficient between brain regions (entries of $R_{ij}$) and the estimate from the point-process analysis (entries of $C_{ij}$). The inset shows the plot for each one of the 1147 subjects. B. Left: performance the point-process based estimation of functional connectivity as a function of the threshold ($\gamma$ in Eq. 2) (mean $\pm$ SD). Elapsed times were obtained for a single subject across 100 repetitions and compared with the performance using linear correlations. Right: Percentage of the original number of data points retained after converting the data to a sparse point-process, plotted as a function of the threshold (for all subjects in the 1000 Functional Connectomes dataset). C. Left: cumulative time required to compute whole-brain voxel-wise connectivity matrices from 1000 subjects extracted from the Functional Connectomes dataset. An un-normalized point-process with $\gamma$ =1 was used. Right: cumulative space required to store 1000 subjects from the Functional Connectomes dataset, both for the full data and for a sparse representation based on a point-process with $\gamma$ =1. }
\end{figure*}

\begin{figure*}[ht]
\centering \includegraphics[width=0.8\textwidth,clip=true,angle=0]{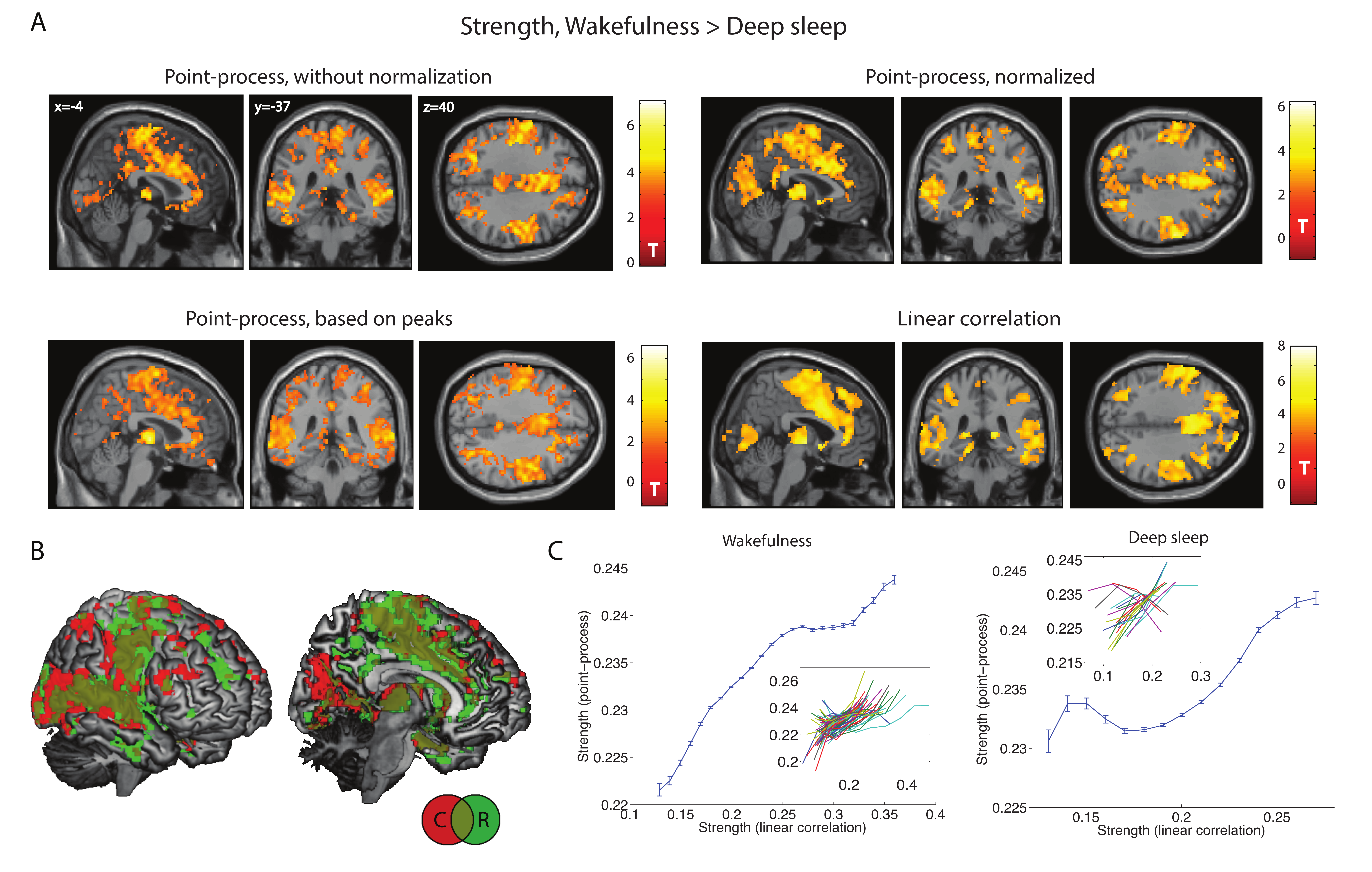} 
\caption{Voxel-wise changes in node strength can be equally observed from $R_{ij}$  and from $\bar{C}_{ij}$. A. Spatial maps showing voxels with decreased strength in deep sleep (N3 sleep) vs. wakefulness, both for the point-process analysis ($C_{ij}$, un-normalized and normalized), for the peak-based point-process and for linear correlations ($R_{ij}$, bottom) (display at p$<$0.05, FWE cluster corrected). B. 3D rendering of the maps in panel A: node strength based on the normalized point-process (red), on linear correlations (green) and their intersection (yellow). C. Plot of the node strength values derived from the point-process vs. those derived from the linear correlation (mean $\pm$ SEM), for wakefulness (left) and for deep sleep (right). Insets show results for individual sleep epochs.}
\end{figure*}

\begin{figure*}[ht]
\centering \includegraphics[width=0.8\textwidth,clip=true,angle=0]{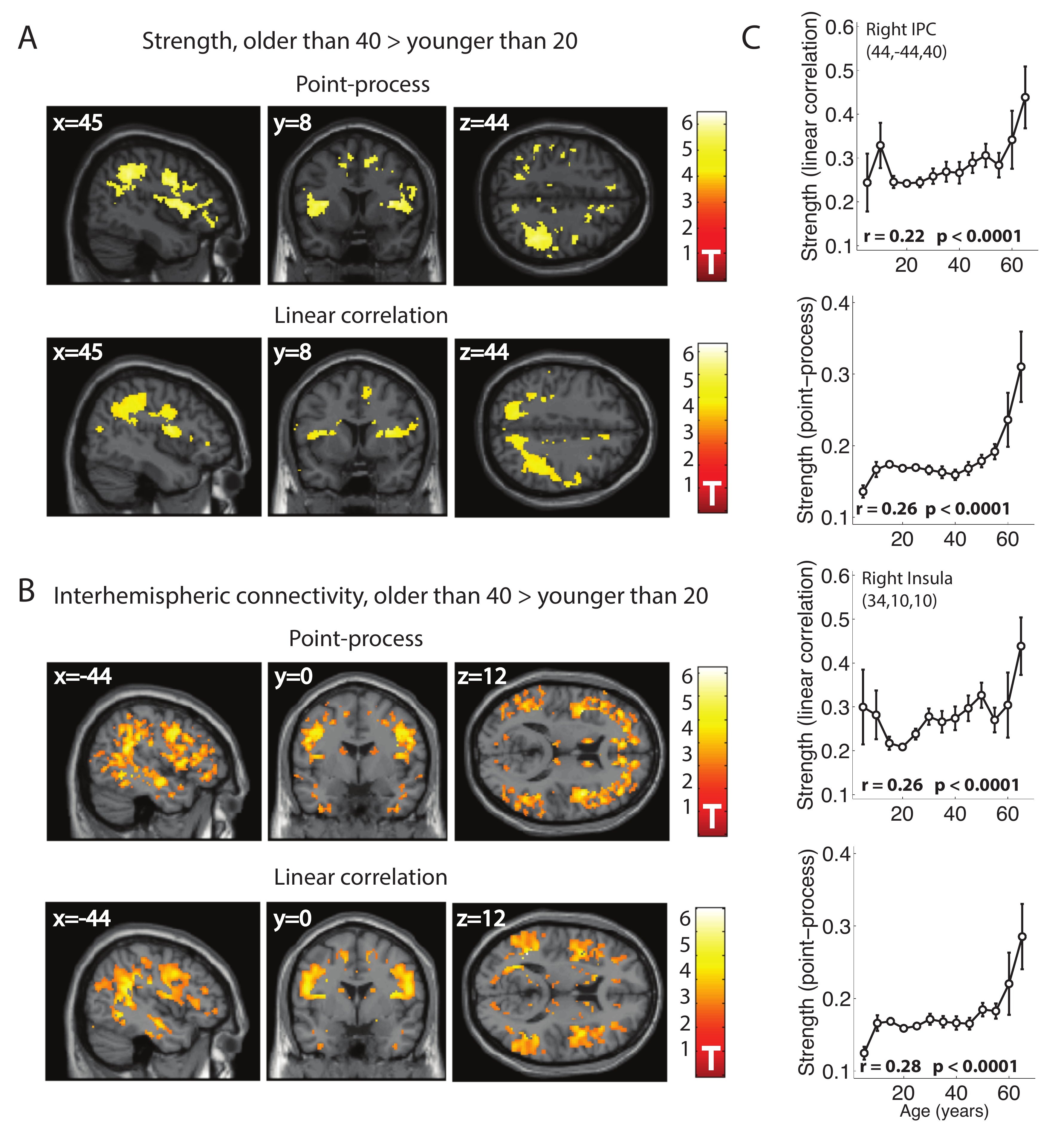} 
\caption{Voxel-wise changes in node strength and interhemispheric connectivity between two age groups ($<$ 20 years and $>$ 40 years) observed from $R_{ij}$  and from $\bar{C}_{ij}$. A. Spatial maps showing voxels with increased strength in the older group when compared to the younger group, both for the normalized point-process ($\bar{C}_{ij}$, top) and for linear correlation ($R_{ij}$, bottom). Only voxels passing a threshold of p $<$ 0.05 (FWE corrected) are shown. B. Spatial maps showing voxels with increased interhemispheric connectivity in the older group when compared with the younger group, both for results obtained from the normalized point-process ($\bar{C}_{ij}$, top) and for linear correlation ($R_{ij}$, bottom). Only clusters passing a threshold of p $<$ 0.05 (FWE corrected) are shown. C. Plots subject of age (in years) vs. strength values (derived from linear correlations and the normalized point-process) extracted from two regions of interest (right Inferior Parietal Cortex - IPC, and right insular cortex) (mean $\pm$ SEM).}
\end{figure*}

\end{document}